\documentclass[twocolumn,epjc3]{svjour3}
\RequirePackage[numbers,sort&compress]{natbib}

\makeatletter
\let\cl@chapter\undefined
\makeatletter

\usepackage[utf8]{inputenc}
\usepackage[pdftex]{graphicx}
\usepackage{float}
\usepackage{amsmath}
\usepackage{amssymb}
\usepackage{mathtools}
\usepackage{siunitx}
\usepackage{amsfonts}
\usepackage{dsfont}
\usepackage{array}
\usepackage{bm}
\usepackage{mathrsfs}
\usepackage{pifont}
\usepackage{multirow}
\usepackage{upgreek}
\usepackage[dvipsnames]{xcolor}
\usepackage[pdftex,
  pdftitle={Testing a new method for scattering in finite volume in the phi to the four theory},
  pdfauthor={M. Garofalo, F. Romero-Lopez, A. Rusetsky, C. Urbach},
  bookmarks,
  colorlinks,
  linkcolor=myblue,
  citecolor=mymagenta,
  menucolor=black,
  urlcolor=myblue,
  plainpages=false,
  pdfpagelabels,
  hypertexnames=false]{hyperref}
\usepackage{cleveref}
\usepackage{verbatim}
\usepackage{slashed}
\usepackage{ucs}
\usepackage{subfigure}

\usepackage{longtable,booktabs,array}

\definecolor{mymagenta}{RGB}{200, 0, 100}
\definecolor{myblue}{RGB}{45, 48, 146}

\graphicspath{{plots/}}

\journalname{Eur. Phys. J. C}
\begin{document}

\title{Testing a new method for scattering in finite volume in the $\phi^4$ theory}
\subtitle{}


\author{Marco Garofalo\thanksref{e1,HISKP}
        \and
        Fernando Romero-L\'opez\thanksref{IFIC}
        \and
        Akaki Rusetsky\thanksref{HISKP,TSU}
        \and\\
        Carsten Urbach\thanksref{HISKP}
}

\thankstext{e1}{e-mail: garofalo@hiskp.uni-bonn.de}


\institute{HISKP (Theory), Rheinische Friedrich-Wilhelms-Universit\"at Bonn, Nussallee 14-16, 53115 Bonn, Germany \label{HISKP}
           \and
           IFIC, CSIC-Universitat de Val\`encia, 46980 Paterna, Spain\label{IFIC}
           \and
           Tbilisi State University, 0186 Tbilisi, Georgia\label{TSU}
}

\date{\today}

\maketitle

\begin{abstract}
We test an alternative proposal by Bruno and Hansen~\cite{Bruno:2020kyl} to extract the scattering length from lattice simulations in a finite volume. For this, we use a scalar $\phi^4$ theory with two mass nondegenerate particles and explore various strategies to implement this new method. We find that the results are comparable to those obtained from the Lüscher method, with somewhat smaller statistical uncertainties at larger volumes. 
\end{abstract}

\section{Introduction}

Lattice QCD has been shown to be a powerful tool to determine scattering quantities from first principles. The standard approach is the Lüscher method~\cite{Luscher:1990ck}, which relates the finite-volume spectrum obtained from the lattice to the infinite-volume scattering amplitude. It has been applied to many physical systems, including results at the physical point---see Ref.~\cite{Briceno:2017max} for a review. The formalism has also been recently extended to three particles with three different but conceptually equivalent formulations available in the literature at present~\cite{Hansen:2014eka,Hansen:2015zga,Hammer:2017uqm,Hammer:2017kms,Mai:2017bge}, see Refs.~\cite{Hansen:2019nir,Mai:2021lwb} for recent reviews.

In Ref.~\cite{Bruno:2020kyl}, the authors propose a new strategy to extract scattering quantities. Henceforth, this will be referred to as the BH method. This approach is based on the usage of four-point functions rather than energy levels. The hope is that this approach can be generalised more easily to multi-hadron processes. 

As pointed out by the authors, the case of threshold kinematics is particularly favourable as it allows for a direct extraction of the scattering length, with the $\pi N$ channel being one concrete example.

In this letter we test this novel approach in a scalar $\phi^4$ theory. Using this theory has proven to be an excellent test bed for novel scattering studies, as shown in Refs.~\cite{Sharpe:2017jej,Romero-Lopez:2018rcb,Romero-Lopez:2020rdq}. In order to mimic the $\pi N$ case, we consider two mass nondegenerate real scalar particles. We explore the necessary techniques, and the optimal approach to use the BH method at threshold. Moreover, we compare to the standard Lüscher approach and find good agreement.

\section{Description of the Model}

The Euclidean model used here is composed by two real scalar fields $\phi_i, i=0,1$ with the Lagrangian
  \begin{align}
  {\cal L}= \sum_{i=0,1} \left( \frac{1}{2} \partial_\mu \phi_i\partial_\mu \phi_i +\frac{1}{2}m_i \phi_i^2 +\lambda_i \phi_i^4\right)
+\mu \phi_0^2 \phi_1^2\, ,
\label{eq:lagrangian} 
 \end{align}
with nondegenerate (bare) masses $m_0<m_1$. The Lagrangian has a $Z_2\otimes Z_2$ symmetry $\phi_0\to-\phi_0\otimes \phi_1\to-\phi_1$, which prevents sectors with even and odd number of particles to mix. 

To study the problem numerically,
we define the theory on a finite hypercubic lattice with lattice spacing $a$ and a volume $T \cdot L^3$, where $T$ denotes the Euclidean time length and $L$ the spatial length. We define the derivatives of the Lagrangian (\cref{eq:lagrangian}) on a finite lattice as the finite differences $\partial_\mu \phi(x)=\frac{1}{a}\,(\phi(x+a\mu)-\phi(x))$.  In addition, periodic boundary conditions are assumed in all directions. The discrete action is given in Ref.~\cite{Romero-Lopez:2018rcb} for the complex scalar theory, but it is trivial to adapt to this case. We set $a=1$ in the following for convenience.

\section{Observables}

In Ref.~\cite{Bruno:2020kyl}, Bruno and Hansen derived a relation between the scattering length $a_0$ and the following combination of Euclidean four-point and two-point correlation functions at the two-particle threshold:
\begin{equation}
    C_4^\mathrm{BH}(t_f,t,t_i) \equiv   \frac{\langle \phi_0(t_f)\phi_1(t)\phi_1(t_i) \phi_0(0)\rangle}
{\langle \phi_0(t_f) \phi_0(0)\rangle \langle \phi_1(t)\phi_1(t_i) \rangle} -1,
\end{equation}
with the time ordering $t_f > t > t_i > 0$. The relation of $C_4^\mathrm{BH}$ to the scattering length reads 
\begin{align}
  \begin{split}
    C_4^\mathrm{BH}(t_f,t,t_i)
    \xrightarrow[t\gg t_i\gg 0]{T\gg t_f\gg t}
    \frac{2}{ L^3}&\bigg[ \pi \frac{a_0 }{\mu_{01} }(t-t_i)
      \\
      - 2a_0^2\sqrt{\frac{2 (t-t_i)}{\mu_{01}} }&+ O\left((t-t_i)^0\right)\bigg]\,,
    \label{eq:BH}  
  \end{split}    
\end{align}
where $\mu_{01}=(M_0 M_1)/(M_0+M_1)$ is the reduced mass. It is defined in terms of the renormalized masses $M_0$ and $M_1$ of the two particles. These masses can be extracted as usual from an exponential fit at large time distances of the two-point correlation functions 
$\langle \phi_{i}(t)\phi_{i}(0) \rangle \approx
   A_{1,i} \left(e^{ - M_{i}  t}  + e^{ - M_{i} (T - t)}\right)$ for $i=0,1$.

\section{Numerical result}

\subsection{BH method}

We generate ensembles using the Metropolis-Hastings algorithm with bare masses $m_0=-4.925$ and $m_1=-4.85$, and for simplicity we choose $\lambda_0=\lambda_1=2\mu=2.5$.  The list of ensembles generated in this work with their corresponding measured values of the masses $M_0$ and $M_1$ are compiled in \cref{tab:full_table}. In this model, as observed in previous investigations of the scalar theory \cite{Romero-Lopez:2020rdq}, we do not see relevant effects of excited states in the two-point correlators, i.e., they are dominated by the ground state from the first time slice.

\begin{figure}
    \centering
    \includegraphics[scale=0.55]{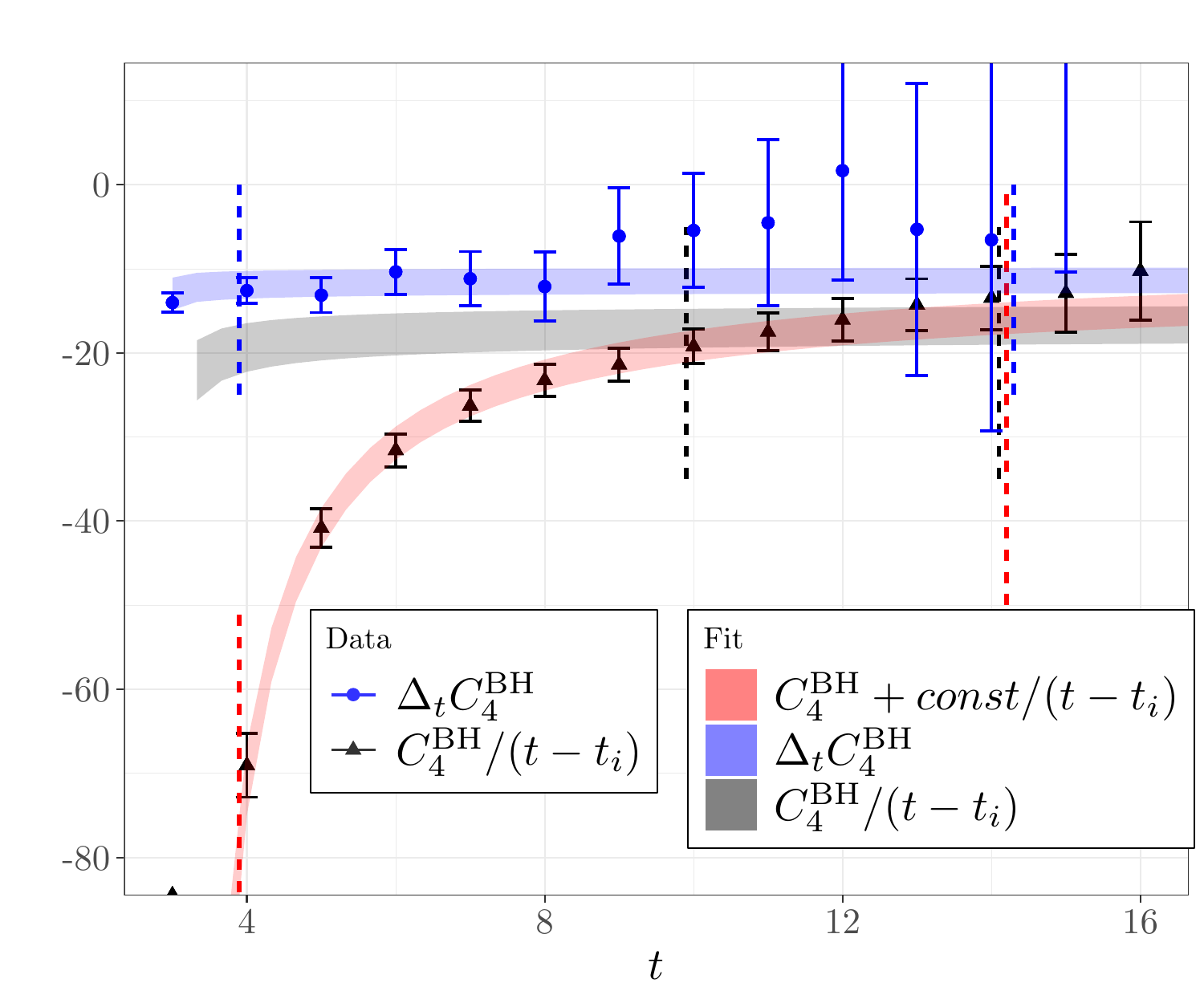}
    \caption{Four-point function of \cref{eq:BH} multiplied by $L^3/2$, for $L=22$ and $T=96$ with $t_i=3$ and $t_f=16$ divided by $(t-t_i)$ black triangles. the dashed vertical lines represent the fit interval, the black band represent the result of the fit \cref{eq:BH} and the red band is the same fit with an extra constant term. The blue circle and band represent the discrete derivative of the correlator \cref{eq:Delta_BH} and the corresponding  fit.}
    \label{fig:BH_03t16}
\end{figure}

In the following we discuss three different strategies to extract the scattering length:
\begin{enumerate}
\item We attempt a direct fit of \cref{eq:BH} the the data. 
\item We include an overall constant in the fit to account for the $O\left((t-t_i)^0\right)$ effect. 
\item We make use of a shifted function at fixed $t_i$ and $t_f$,
$\Delta_t C_4^\mathrm{BH}(t_f,t,t_i)= C_4^\mathrm{BH}(t_f,t+1,t_i)-C_4^\mathrm{BH}(t_f,t,t_i)$,
which cancels the constant term. We then determine $a_0$ by fitting to
  \begin{equation}
    \label{eq:Delta_BH}
    \begin{split}
      \Delta_t C_4^\mathrm{BH}(t_f,t,t_i&) \approx \frac{2}{L^3}\Big[\pi\frac{a_0}{\mu_{01}}\Big. \\
        \Big.-2a_0^2&\sqrt{\frac{2}{\mu_{01}}}\Big(\sqrt{t+1-t_i} - \sqrt{t-t_i}\,\Big)\Big].
    \end{split}
  \end{equation}
\end{enumerate}
The three methods are compared in \cref{fig:BH_03t16} for one of our ensembles. The black triangles represent the correlator of \cref{eq:BH} divided by $(t-t_i)$ with $t_i=3$ and $t_f=16$. This representation is convenient, as it converges towards a constant when $(t-t_i) \to \infty$. From the monotonically increase of the data points, it is clear that the effect of the $\left((t-t_i)^0\right)$ term in \cref{eq:BH} is still sizeable even at large time separations. A fit in the time region $[10,14]$---the black band---is reasonable ($\chi^2/d.o.f\sim0.7$) but results in large uncertainties. The quality of the fit deteriorates very quickly if the fit range is extended: a fit in the time region $[6,14]$ yields a $\chi^2/\mathrm{dof}\sim 5$.

With the second strategy---the red band in \cref{fig:BH_03t16}---one is able to start fitting at significantly smaller $t$-values.
The data are well described with a $\chi^2/\mathrm{dof}\sim 0.2$ 

For the third approach, we study $\Delta_t C_4^\mathrm{BH}(t)$. This is shown in \cref{fig:BH_03t16} as blue circles, and the blue band represents the best fit result with error.
The main advantage of the last strategy is that it allows us to extract the physical information at smaller $t$ without introducing extra parameters in the fit. 
Indeed, the data looks almost constant over the complete $t$-range available.
Only very close to $t_i$ the square root term might become visible.

\begin{figure}
    \centering
    \includegraphics[scale=0.55]{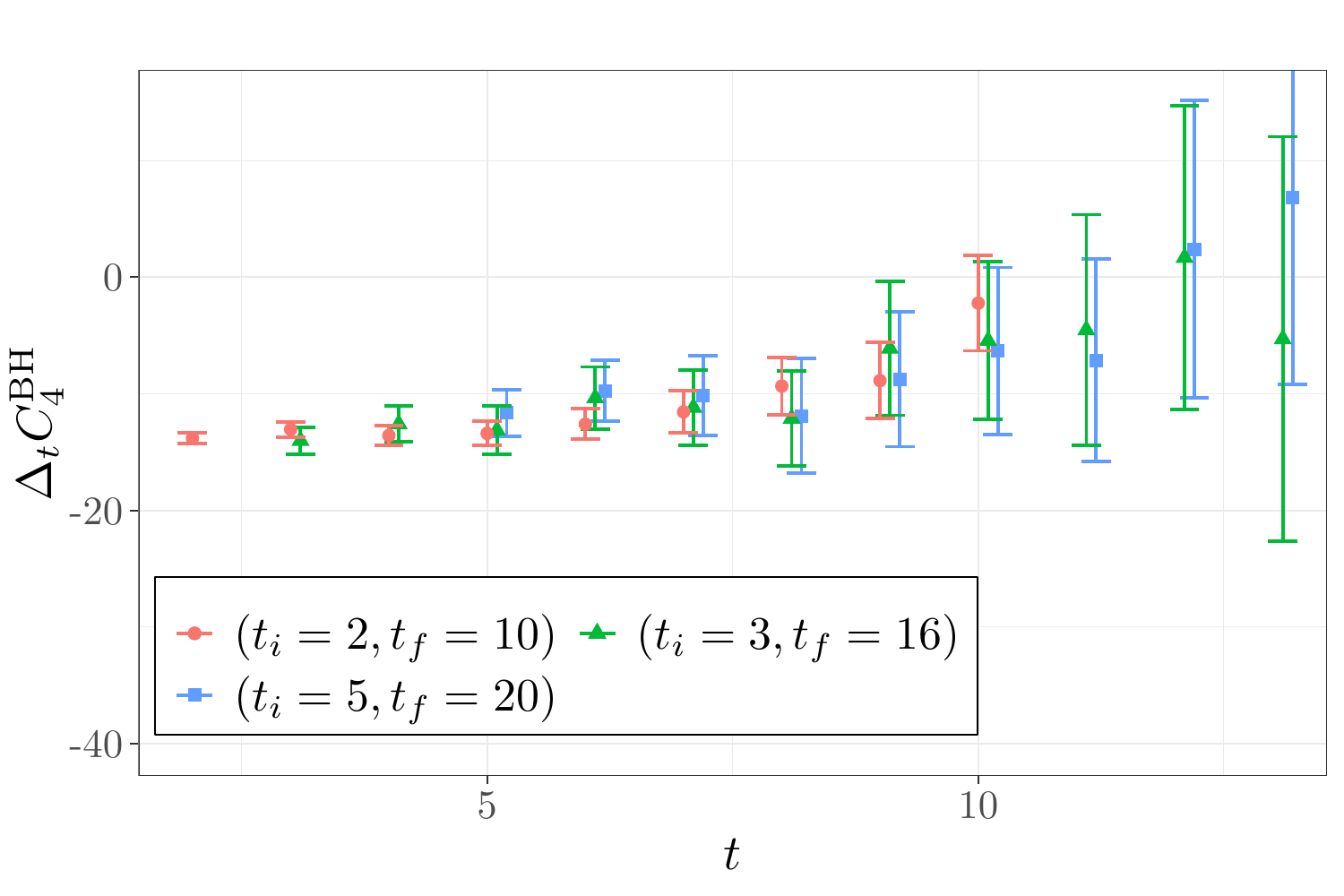}
    \caption{Plot of the discrete derivative of the correlator \cref{eq:Delta_BH}
    for different values of $t_i$ and $t_f$. We do not observe any systematic shift and all correlators 
    are compatible. The points with smaller $t_i$ and $t_f$
    tend to have smaller error.}
    \label{fig:BH_ti}
\end{figure}

For this third strategy, which looks most promising from a systematic point of view, we also investigate the dependence on the choice for $t_{i}$ and $t_{f}$. This is shown in \cref{fig:BH_ti} for the same ensemble as in \cref{fig:BH_03t16}.
We do not observe any significant systematic effect stemming from excited state contributions when changing $t_i$ or $t_f$.
However, we clearly see significantly smaller statistical uncertainties with smaller $t_i$ and $t_f$ values.

\subsection{Comparison to the Lüscher method}

In this section we compare the BH method described above with the Lüscher threshold expansion~\cite{Luscher:1985dn}. The latter relates the two-particle energy shift, defined as $\Delta E_{2}=E_{2}-M_0-M_1$, to the scattering length $a_0$ via
\begin{equation}
  \Delta E_{2} =-\frac{2\pi a_{0}}{\mu_{01} L^3}\left[ 1 + c_1  \frac{a_{0}}{L} + c_2\left(\frac{a_{0}}{L}\right)^2  \right] +O\left(L^{-6}\right)\,,
  \label{eq:luescher_a0}
\end{equation}
with $c_1=-2.837297$, $c_2= 6.375183$ and $E_2$ being the interacting two-particle energy.
$E_2$ can be extracted from 
$C_2(t) = \langle \phi_1(t)\phi_0(t)  \phi_1(0)\phi_0(0) \rangle$, whose large-$t$ behaviour is
\begin{align}
    \begin{split}
         C_2(t)     &\xrightarrow[T-t\gg0]{t\gg0}     A_2 e^{-E_2 \frac{T}{2}} \cosh{\left(E_2 (t-\frac{T}{2})\right)} \\
     +&B_2  e^{-(M_0+M_1) \frac{T}{2}} \cosh{\left((M_1-M_0) (t-\frac{T}{2})\right)}\,.     \label{eq:E2_01}
    \end{split}
\end{align}
Note that the last term is a known thermal pollution due to finite $T$ in the presence of periodic boundary conditions.
Using $M_0$ and $M_1$ as input determined from the corresponding two-point functions, the only additional parameter is $B_2$.

Alternatively, it is possible to eliminate the second term defining
\begin{equation}
\tilde C_2(t)=C_2(t)/\cosh{\left((M_1-M_0) (t-\frac{T}{2})\right)},
\end{equation}
and then taking the finite derivative
\begin{align}
    \begin{split}
  \Delta_t\tilde C_2(t)&=\tilde C_2(t+1)-\tilde C_2(t)  \\
   & \propto \frac{ \sinh{E_2 \left( t-\frac{T}{2}+\frac{1}{2}\right)}}{\sinh{ (M_1-M_0) \left(t-\frac{T}{2} + \frac{1}{2}\right)}}.
    \label{eq:Delta_E2_01}
    \end{split}
\end{align}
The two-particle energies obtained from \cref{eq:E2_01} are compatible to those from \cref{eq:Delta_E2_01}. The results are reported in \cref{tab:full_table}, along with the values for 
 the scattering length $a_0$ computed from $E_2$ using \cref{eq:luescher_a0}.

\begin{figure}
    \centering
    \includegraphics[scale=0.55]{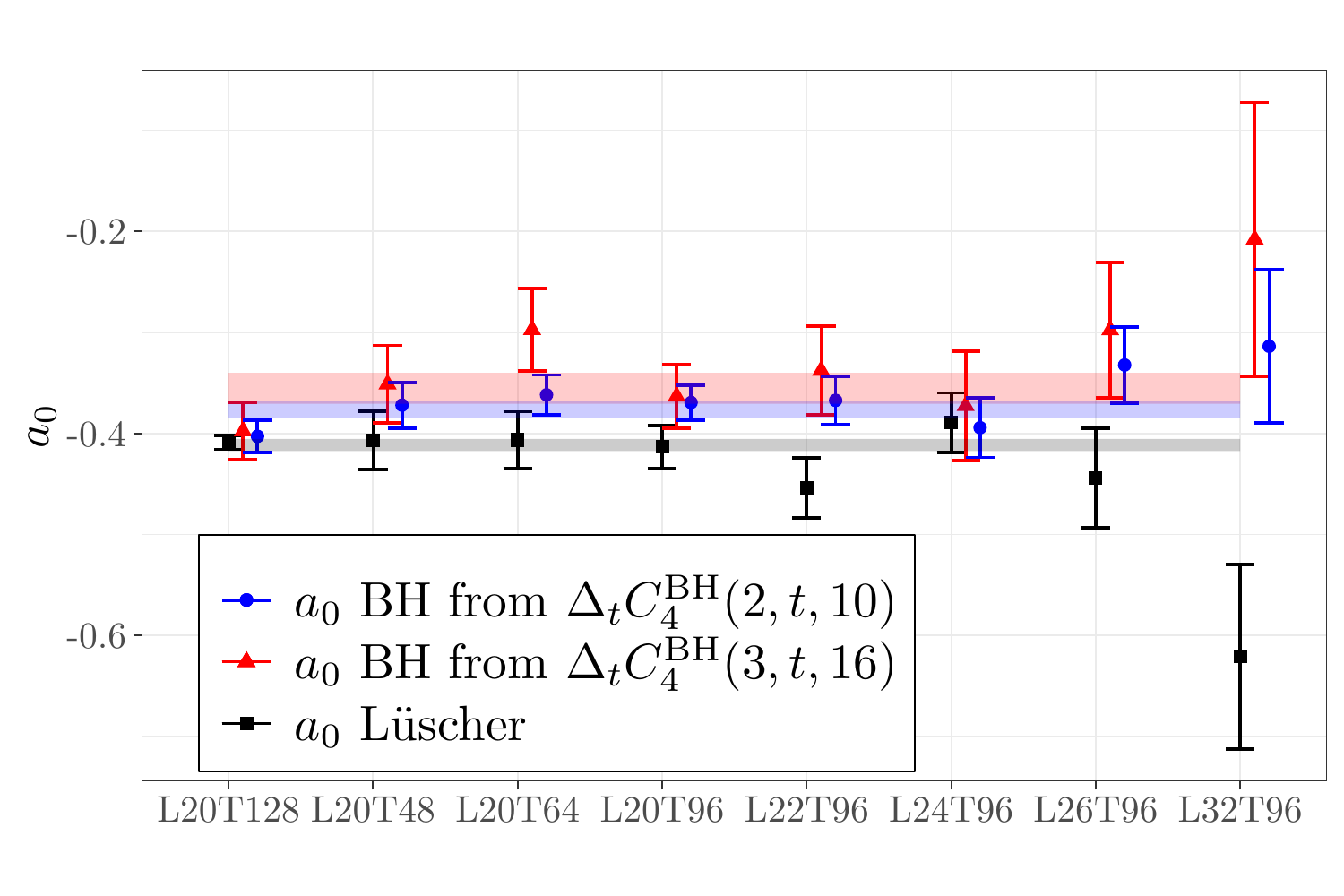}
    \caption{Comparison of $a_0$ computed with  BH method \cref{eq:Delta_BH} with $t_i=2$ and $t_f=10$ (blue circles), with $t_i=3$ and $t_f=16$ (red triangles) and  Lüscher method  \cref{eq:luescher_a0} (black squares). The horizontal bands correspond to the weighted average of each method. }
    \label{fig:compare_L_BH}
\end{figure}

 The comparison between the BH and the Lüscher method is depicted in \cref{fig:compare_L_BH} for all our ensembles. The values are compatible with each other, however the BH method
 gives systematically larger values for $a_0$.

\begin{table*}
{\footnotesize
    \centering
    \setlength{\tabcolsep}{3pt}
    
   \setlength{\tabcolsep}{3pt}
    \begin{tabular}{cc|cc|cc|cc|ccc}
\hline\hline
T & L & \(M_0\) & \(M_1\) &\multicolumn{2}{c|}{$E_2$} &\multicolumn{2}{c|}{$a_0$ Lüscher} & \multicolumn{3}{c}{$a_0$ BH}    \\
\hline
 &  &  &  & \(  C_2\) & \( \Delta_t\tilde C_2\) & \(  C_2\) & \( \Delta_t\tilde C_2\) & \(\Delta_t C_4^\mathrm{BH}\)(3,t,16) 
& \(C_4^\mathrm{BH}+c\)  & \(C_4^\mathrm{BH}\)(2,t,10)  \\
\hline
48 & 20 & 0.14675(5) & 0.27487(5) & 0.4252(3) & 0.4253(3) &
-0.41(3) & -0.42(3) & -0.35(4) & -0.35(6) & -0.37(2) \\
64 & 20 & 0.14659(5) & 0.27480(5) & 0.4249(3) & 0.4250(3) &
-0.41(3) & -0.41(4) & -0.30(4) & -0.29(6) & -0.38(2) \\
96 & 20 & 0.14662(4) & 0.27487(4) & 0.4251(2) & 0.4251(3) &
-0.41(2) & -0.41(3) & -0.36(3) & -0.36(4) & -0.38(1) \\
96 & 22 & 0.14604(3) & 0.27470(4) & 0.4237(2) & 0.4237(3) &
-0.45(3) & -0.45(5) & -0.34(4) & -0.31(6) & -0.37(2) \\
96 & 24 & 0.14574(4) & 0.27458(4) & 0.4223(2) & 0.4221(3) &
-0.39(3) & -0.36(6) & -0.36(5) & -0.41(7) & -0.39(2) \\
96 & 26 & 0.14547(4) & 0.27455(3) & 0.4218(2) & 0.4219(3) &
-0.44(5) & -0.47(8) & -0.30(7) & -0.3(1) & -0.36(3) \\
96 & 32 & 0.14521(4) & 0.27449(4) & 0.4210(2) & 0.4213(3) &
-0.62(9) & -0.7(1) & -0.2(1) & -0.1(2) & -0.35(5) \\
128 & 20 & 0.14668(3) & 0.27484(3) & 0.42509(7) & 0.4251(3) &
-0.409(7) & -0.41(3) & -0.40(3) & -0.39(3) & -0.40(1) \\
\hline\hline
\end{tabular}
    \caption{Values of $a_0$, $M_0$, $M_1$ and $E_2$ measured.
    The column \(\Delta_t C_\mathrm{BH}\) corresponds to the value of $a_0$ fitted with \cref{eq:Delta_BH} fixing $t_i=3$ and $t_f=16$ or $t_i=2$ and $t_f=10$,
     the column \(C_\mathrm{BH}+c\) is the result of the fit \cref{eq:BH} adding a constant term. The two-particle energy 
     $E_2$ is computed form $C_2$ with the fit of \cref{eq:E2_01} and from $\Delta \tilde C_2$ with 
     \cref{eq:Delta_E2_01}. The corresponding value of $a_0$ computed with the Lüscher method is reported in the corresponding columns.
    }
    \label{tab:full_table}
}
\end{table*}

\section{Conclusion}

In this letter, we have investigated the BH method, proposed in Ref.~\cite{Bruno:2020kyl}, using a scalar theory on the lattice. We have indeed verified that it is a viable method to obtain the scattering length, and that it produces results that are compatible with those of the Lüscher method~\cite{Luscher:1985dn}.
The most reliable strategy to analyse the four-point function is found to be the use of finite differences in time to remove an overall constant term.

We observe a systematic difference between the Lüscher and BH method. Interestingly, for each ensemble separately both determinations appear compatible. The systematic trend becomes evident only after averaging over all runs, as shown in the bands of fig.~\ref{fig:compare_L_BH}. This might be attributed to different lattice artefacts, since both methods represent different estimators for $a_0$. We are not able to check this hypothesis here, because we cannot take the continuum limit. However, the different systematics of the two methods offer in general a useful opportunity for cross checks.

The statistical error is similar in both approaches. Also the scaling in $L$ appears to be similar, with maybe a slight advantage for the BH method. However, any advantage of one method compared to another one will in general depend on the theory considered. We conclude that it seems promising to use the BH method in lattice QCD for instance for $\pi N$ scattering, where also the lattice spacing dependence could be investigated.

\begin{acknowledgements}
  We gratefully acknowledge helpful discussions with with M.~Bruno and M.~T.~Hansen.
  FRL acknowledges financial support from the Generalitat Valenciana grants PROMETEO/2019/083 and CIDEGENT/2019/040, the EU project H2020-MSCA-ITN-2019//860881-HIDDeN, and the Spanish project FPA2017-85985-P.
  This work
  is supported in part by the Deutsche Forschungsgemeinschaft (DFG,
  German Research Foundation) and the 
  NSFC through the funds provided to the Sino-German
  Collaborative Research Center CRC 110 “Symmetries
  and the Emergence of Structure in QCD” (DFG Project-ID 196253076 - TRR 110, NSFC Grant No. 12070131001).
  AR acknowledges support from Volkswagenstiftung (Grant No. 93562) and the Chinese Academy of Sciences (CAS) President’s International Fellowship Initiative (PIFI) (Grant No. 2021VMB0007).
  The C\texttt{++} Performance Portability Programming Model Kokkos \cite{CarterEdwards20143202} and
  the open source software packages R~\cite{R:2019} have 
  been used.
  We thank B.~Kostrzewa for useful discussions on Kokkos.
\end{acknowledgements}

\bibliographystyle{spphys} 
\bibliography{bibliography}

\begin{thebibliography}{10}
\providecommand{\url}[1]{{#1}}
\providecommand{\urlprefix}{URL }
\expandafter\ifx\csname urlstyle\endcsname\relax
  \providecommand{\doi}[1]{DOI \discretionary{}{}{}#1}\else
  \providecommand{\doi}{DOI \discretionary{}{}{}\begingroup
  \urlstyle{rm}\Url}\fi

\bibitem{Bruno:2020kyl}
M.~Bruno, M.T. Hansen, Journal of High Energy Physics \textbf{2021}(6) (2021).
\newblock \urlprefix\url{http://dx.doi.org/10.1007/JHEP06(2021)043}

\bibitem{Luscher:1990ck}
M.~Lüscher, U.~Wolff, Nucl. Phys. \textbf{B339}, 222 (1990)

\bibitem{Briceno:2017max}
R.A. Briceno, J.J. Dudek, R.D. Young, Rev. Mod. Phys. \textbf{90}(2), 025001
  (2018)

\bibitem{Hansen:2014eka}
M.T. Hansen, S.R. Sharpe, Phys. Rev. D \textbf{90}(11), 116003 (2014)

\bibitem{Hansen:2015zga}
M.T. Hansen, S.R. Sharpe, Phys. Rev. D \textbf{92}(11), 114509 (2015)

\bibitem{Hammer:2017uqm}
H.W. Hammer, J.Y. Pang, A.~Rusetsky, JHEP \textbf{09}, 109 (2017)

\bibitem{Hammer:2017kms}
H.W. Hammer, J.Y. Pang, A.~Rusetsky, JHEP \textbf{10}, 115 (2017)

\bibitem{Mai:2017bge}
M.~Mai, M.~D\"oring, Eur. Phys. J. A \textbf{53}(12), 240 (2017)

\bibitem{Hansen:2019nir}
M.T. Hansen, S.R. Sharpe, Ann. Rev. Nucl. Part. Sci. \textbf{69}, 65 (2019)

\bibitem{Mai:2021lwb}
M.~Mai, M.~Döring, A.~Rusetsky, EPJ ST  (2021)

\bibitem{Sharpe:2017jej}
S.R. Sharpe, Phys. Rev. D \textbf{96}(5), 054515 (2017).
\newblock [Erratum: Phys.Rev.D 98, 099901 (2018)]

\bibitem{Romero-Lopez:2018rcb}
F.~Romero-L\'opez, A.~Rusetsky, C.~Urbach, Eur. Phys. J. C \textbf{78}(10), 846
  (2018)

\bibitem{Romero-Lopez:2020rdq}
F.~Romero-L\'opez, A.~Rusetsky, N.~Schlage, C.~Urbach, JHEP \textbf{02}, 060
  (2021)

\bibitem{Luscher:1985dn}
M.~Lüscher, Commun.Math.Phys. \textbf{104}, 177 (1986).
\newblock \urlprefix\url{http://inspirehep.net/record/222569}

\bibitem{CarterEdwards20143202}
H.C. Edwards, C.R. Trott, D.~Sunderland, Journal of Parallel and Distributed
  Computing \textbf{74}(12), 3202  (2014).
\newblock
  \urlprefix\url{http://www.sciencedirect.com/science/article/pii/S0743731514001257}

\bibitem{R:2019}
{R Core Team}, \emph{R: A Language and Environment for Statistical Computing}.
\newblock R Foundation for Statistical Computing, Vienna, Austria (2019).
\newblock \urlprefix\url{https://www.R-project.org/}

\end{thebibliography}

\end{document}